\documentclass[aps,prd,superscriptaddress,groupedaddress,nofootinbib,nobibnotes]{revtex4}

\usepackage{graphicx}
\usepackage{dcolumn}
\usepackage{bm}
\usepackage{amssymb}
\usepackage{epstopdf}
\usepackage{amsmath}
\usepackage{amsfonts}
\usepackage{color}
\usepackage{mathrsfs}
\usepackage[normalem]{ulem}

\newcommand{\be}{\begin{equation}}
\newcommand{\ee}{\end{equation}}
\newcommand{\ba}{\begin{eqnarray}}
\newcommand{\ea}{\end{eqnarray}}
\newcommand{\nn}{\nonumber}
\newcommand{\barr}{\begin{array}}
\newcommand{\earr}{\end{array}}

\newcommand\lsim{\mathrel{\rlap{\lower4pt\hbox{\hskip1pt$\sim$}}
        \raise1pt\hbox{$<$}}}
\newcommand\gsim{\mathrel{\rlap{\lower4pt\hbox{\hskip1pt$\sim$}}
        \raise1pt\hbox{$>$}}}

\def\k{{\bf k}}

\def\n{{\bf n}}

\def\fnl{f_{NL}}

\newcommand{\rmn}{\mathrm}
\newcommand{\mbf}[1]{\mathbf{#1}}

\begin{document}

\title{Constraining local non-Gaussianities with kSZ tomography}

\author{Moritz M\"unchmeyer}
\affiliation{Perimeter Institute for Theoretical Physics, Waterloo, ON N2L 2Y5, Canada}

\author{Mathew~S.~Madhavacheril}
\affiliation{Princeton University, Department of Astrophysical Sciences, Princeton NJ 08540, USA}

\author{Simone Ferraro}
\affiliation{Berkeley Center for Cosmological Physics, University of California, Berkeley CA 94720, USA}
\affiliation{Miller Institute for Basic Research in Science, University of California, Berkeley CA 94720, USA}

\author{Matthew~C.~Johnson}
\affiliation{Perimeter Institute for Theoretical Physics, Waterloo, ON N2L 2Y5, Canada}
\affiliation{Department of Physics and Astronomy, York University, Toronto, Ontario, M3J 1P3, Canada}

\author{Kendrick M. Smith}
\affiliation{Perimeter Institute for Theoretical Physics, Waterloo, ON N2L 2Y5, Canada}

\date{\today}

\begin{abstract}
Kinetic Sunyaev Zel'dovich (kSZ) tomography provides a powerful probe of the radial velocity field of matter in the Universe. By cross-correlating a high resolution CMB experiment like CMB S4 and a galaxy survey like DESI or LSST, one can measure the radial velocity field with very high signal to noise over a large volume of the universe. In this paper we show how this measurement can be used to improve constraints on primordial non-Gaussianities of the local type. The velocity field provides a measurement of the unbiased matter perturbations on large scales, which can be cross-correlated with the biased large-scale galaxy density field. This results in sample variance cancellation for a measurement of scale-dependent bias due to a non-zero $\fnl$. Using this method we forecast that CMB S4 and LSST combined reach a sensitivity $\sigma_{\fnl} \sim 0.5$, which is a factor of three improvement over the sensitivity using LSST alone (without internal sample variance cancellation). We take into account critical systematics like photometric redshifts, the kSZ optical depth degeneracy, and systematics affecting the shape of the galaxy auto-power spectrum and find that these have negligible impact, thus making kSZ tomography a robust probe for primordial non-Gaussianities. We also forecast the impact of mass binning on our constraints. The techniques proposed in this paper could be an important component of achieving the theoretically important threshold of $\sigma_{\fnl} \lesssim 1$ with future surveys.

\end{abstract}

\maketitle


\section{Introduction}

Detecting or tightly constraining primordial non-Gaussianity is one of the main goals of many upcoming large-scale structure surveys and CMB experiments. A detection would provide invaluable information about interactions in the inflationary universe, probing physics at ultra high energy scales that are not otherwise accessible to experiments. Primordial non-Gaussianities have been classified extensively according to production mechanisms, symmetries and field content. A particularly simple and important class are so called local type non-Gaussianities, which are produced generically in the presence of more than one light degree of freedom during inflation (multi-field inflation). A well-known threshold is that multi-field inflation predicts $\fnl \gtrsim 1$ (see e.g.~\cite{Alvarez:2014vva}). For comparison, the current best bound is $\fnl = 2.5 \pm 5.7$, coming from the latest Planck satellite CMB analysis~\cite{Ade:2015ava}. Unfortunately the constraint from the primary CMB is already close to being saturated, as a large fraction of the available modes in temperature and polarization have been measured. 

To reach the multifield threshold, a three-dimensional probe of the universe is needed, which contains many more modes than the two-dimensional CMB sky. Non-Gaussianities can in principle be measured with large-scale structure surveys by measuring the bispectrum of the galaxy distribution. Unfortunately, non-linear evolution by gravity induces large bispectra which are hard to disentangle from those expected from primordial physics. However in the case of local non-Gaussianities, it is possible to obtain excellent constraints from the power spectrum alone, through a measurement of the scale-dependence of galaxy bias~\cite{Dalal:2007cu}. On large scales, the presence of $\fnl \neq 0$ induces a bias relating the matter and galaxy distribution which scales as $1/k^2$, providing a rather unique signal that is not mimicked by changes of the standard cosmological parameters. This signal is the main path through which upcoming galaxy surveys hope to improve constraints on $\fnl$. Reaching the multi-field threshold $\fnl = 1$ from galaxy surveys remains difficult, due to the limited number of large-scale modes within the survey volume, i.e. due to cosmic variance. However, if one can measure tracers of different bias (or biased and unbiased tracers), these can be compared to determine the scale-dependent bias induced by non-Gaussianities, without cosmic variance~\cite{Seljak:2008xr}. This idea, known as sample variance cancellation, can be achieved either by measuring different sets of galaxies, or by cross-correlating the galaxy distribution with an independent probe of large-scale structure not based on galaxies. In particular, recently Schmittful and Seljak proposed to use the CMB lensing potential as a probe of the unbiased matter distribution and to cross-correlate it with a galaxy survey to measure $\fnl $ through sample variance cancellation~\cite{Schmittfull:2017ffw}. Decisive for the power of sample variance cancellation is the correlation coefficient between the biased and unbiased modes, which is challenging in the case of CMB lensing because of the very broad lensing kernel. 

Here we present a new method to improve $\fnl$ measurements by using kinetic Sunyaev Zel'dovich (kSZ) tomography~\cite{Ho09,Shao11b, Zhang11b, Zhang01,Munshi:2015anr,2016PhRvD..93h2002S,Ferraro:2016ymw,Hill:2016dta,Zhang10d,Zhang:2015uta,Terrana2016} to measure the radial velocity field of matter in the universe. The kSZ effect~\cite{Sunyaev1980} is a secondary CMB temperature anisotropy induced by the scattering of CMB photons from the bulk-motion of free-electrons in the post-reionization Universe. This effect is the dominant blackbody contribution to the CMB on small scales (at $\ell \agt 4000$), and will be measured at high significance by future experiments. The direct cross-correlation of a high-resolution CMB map with a galaxy survey can be used to reconstruct the radial velocity field in a {\em 3-dimensional volume}, thus providing an additional probe of large scale structure~\cite{Zhang10d,Terrana2016,Deutsch:2017ybc}. Because the radial velocity field is an unbiased tracer, it can be combined with a galaxy survey to realize the idea of sample variance cancelation. In our approach, sample variance cancellation is particularly powerful, due to a very good correlation coefficient of the reconstructed velocity field and the galaxy distribution. This is possible due to the very low noise in the large-scale velocity field reconstruction. 

We forecast the improvement of $\sigma_{\fnl}$ due to the inclusion of kSZ tomography data for two baseline experimental configurations, `baseline 1' corresponding to DESI~\cite{Aghamousa:2016zmz} + a CMB experiment similar to Simons Observatory, and `baseline 2' corresponding to LSST~\cite{LSSTScienceCollaboration2009} + CMB S4~\cite{CMBS42016}. Our forecast is based on the kSZ tomography bispectrum formalism developed in~\cite{kszbispectrumpaper}, which allows us to make realistic forecasts including photo-z errors, kSZ optical depth degeneracy and redshift space distortions. We find, depending on the redshift range included in the analysis, that improvement factors on $\sigma_{\fnl}$ in the range $\sim 2-10$ are possible by including kSZ tomography. If sample variance cancellation can be realized within the galaxy sample itself, e.g. by considering populations with different halo mass~\cite{2011PhRvD..84h3509H}, improvement factors are more modest for measurable halo masses, but can still be a factor of ~2. Because future surveys are only just on the cusp of attaining $\sigma_{\fnl}=1$, an improvement factor of a few could yield a significant detection or provide a constraint that significantly trims the allowed regions of model space, for example severely constraining~\cite{Alvarez:2014vva} curvaton scenarios~\cite{Linde:1996gt,Enqvist:2001zp,Lyth:2001nq,Moroi:2001ct}. 

The paper is organized as follows. In Sec.~\ref{sec:measuringfnl} we recall scale dependent bias and explain how kSZ tomography can be used for sample variance cancellation. In Sec.~\ref{sec:experiments} we describe the experimental parameters in our forecast. The Fisher forecast setup is described in Sec.~\ref{sec:fishersetup} and its results are discussed in Sec.~\ref{sec:fisherresults}. We summarize our results in Sec.~\ref{sec:conclusions}.

\section{Measuring $f_{NL}$  with kSZ tomography}
\label{sec:measuringfnl}

To explain in detail how kSZ tomography can be used to constrain local non-Gaussianities, we first recall the standard results on scale-dependent bias and then illustrate why kSZ tomography is very well suited for measuring $\fnl$ using sample variance cancellation.

\subsection{Scale dependent bias}

In the presence of primordial non-Gaussianity, a biased tracer acquires a scale-dependent bias proportional to $f_{NL}$. In particular, on large scales, the matter-halo and halo-halo power spectra can be written as~\cite{Dalal:2007cu}
\ba
P_{mh}(k, z) &=& \left( b_h + \fnl \frac{\beta_f}{\alpha(k,z)} \right) P_{mm}(k,z) \\
P_{hh}(k, z) &=& \left( b_h + \fnl \frac{\beta_f}{\alpha(k,z)} \right)^2 P_{mm}(k,z)  \label{eq:ng_bias1} 
\ea
Here we have defined (see for example~\cite{Ferraro:2014jba}) 
\be
\alpha(k,z) = \frac{2 k^2 T(k)}{3 \Omega_m H_0^2} D(z)
\ee
so that the matter overdensity $\delta_m(\k,z)$ is related to the primordial potential $\Phi(\k)$ through the Poisson equation as
\be
\delta_m(\k,z) = \alpha(k,z) \Phi(\k) \ .
\ee
The linear growth function $D(z)$ is normalized so that $D(z) = 1 / (1+z)$ during matter domination and $T(k)$ is the transfer function normalized to 1 at low $k$. The quantity $b_h$ is the Eulerian halo bias and depends on the halo mass. 
The non-Gaussian bias parameter $\beta_f$ is well approximated by
\be
\beta_f = 2 \delta_c (b_h-1).
\ee
We will take $\delta_c = 1.42$, as appropriate for the Sheth-Tormen halo mass function. 

Here and below we write large-scale halo power spectra, which involve the halo bias, with a subscript ``h'', while we write small scale galaxy power spectra that appear in kSZ tomography with a subscript ``g''. In the forecast we will however assume abundance matching for the large-scale halo power spectra, i.e. each halo is assumed to be populated by a single galaxy in its center. This facilitates forecasting with experimental galaxy number densities and does not strongly affect the results. Under this assumption the words halo and galaxy can be used interchangeably. On the other hand the small scale galaxy power spectra will be calculated within the halo model including the halo occupation distribution (HOD)~\cite{2012ApJ...744..159L,Leauthaud:2011zt} and therefore satellite galaxies. The calculation of halo model powerspectra is reviewed in App.~\ref{app:halomodel}.

As shown in~\cite{Deutsch:2017ybc,kszbispectrumpaper}, the noise on the kSZ tomography reconstruction of the radial velocity field is given by
\be
N_{vv}^{\rm rec}(k_L,\mu) = \mu^{-2} \frac{2\pi \chi_*^2}{K_*^2} 
  \left[ \int dk_S \, k_S \left( \frac{P_{ge}(k_S)^2}{P_{gg}^{\rm tot}(k_S) C_l^{\rm tot}} \right)_{l=k_S\chi_*} \right]^{-1}  \label{eq:Nvv}
\ee
where we approximated the observed part of the universe by a box at distance $\chi_*$. The details this box approximation are explained in~\cite{kszbispectrumpaper}. Here and below the star suffix indicates the center of the box. $P_{ge}$ is the small-scale galaxy-electron cross power spectrum and $P_{gg}$ is the small-scale galaxy auto power spectrum, which are both dominated by the 1-halo term  on the scales of interest \cite{kszbispectrumpaper}. The kSZ radial weight function $K_*$ at the corresponding redshift $z_*$ is given by $K(z) = -T_{\rm CMB} \sigma_T n_{e,0} x_e e^{-\tau(z)} (1+z)^2$, where $\sigma_T$ is the Thomson cross section, $n_{e,0}$ is the comoving electron density, $x_e(z)$ is the ionized fraction and $\tau$ is the optical depth.
 
\begin{figure}[tbh]
  \includegraphics[width=0.45\textwidth]{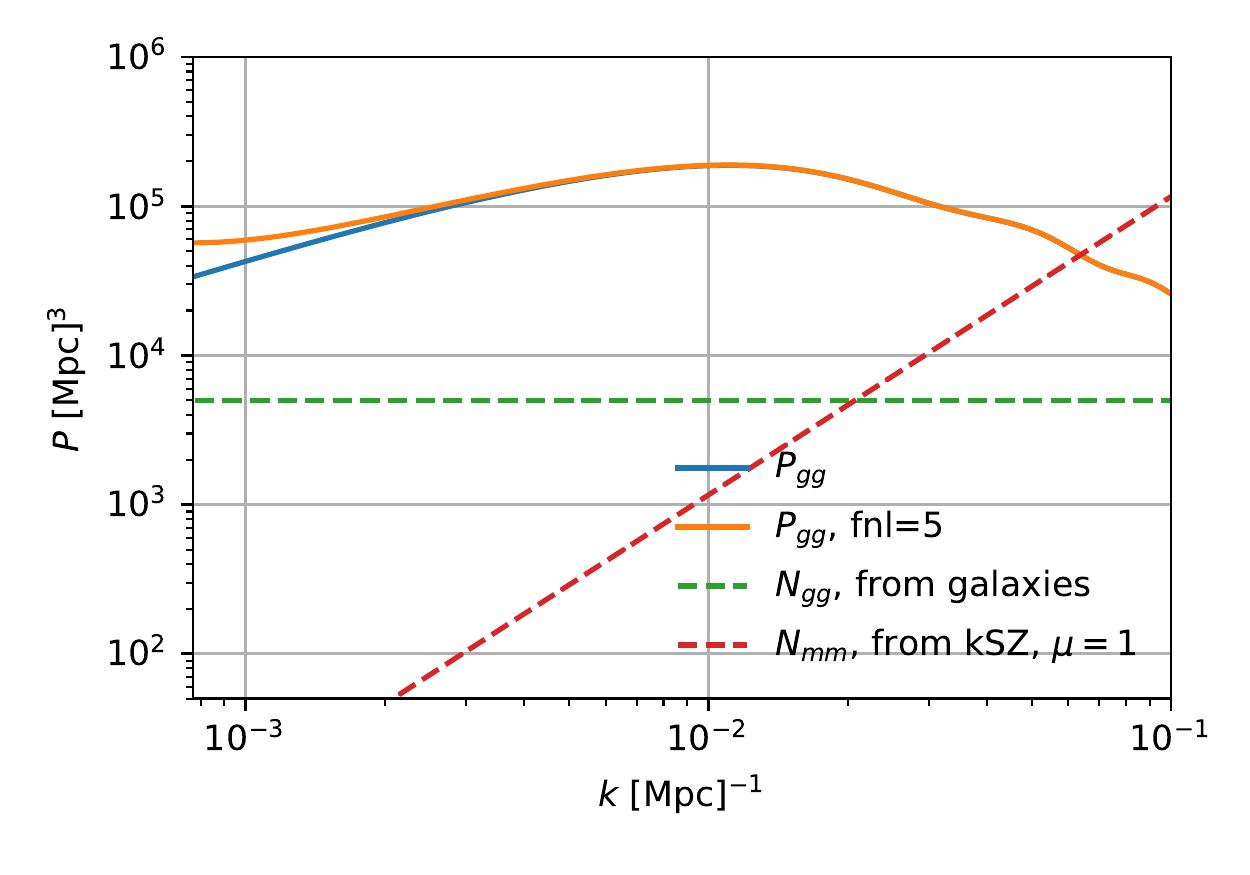}
  \caption{Galaxy power spectra and noise power spectra for an experimental configuration corresponding to the DESI experiment (with galaxy density and CMB noise according to our baseline 1 defined in Table~\ref{tab:baseline}). At large scales, where the $\fnl$ signal is strongest, we get a much lower noise using kSZ tomography than with galaxies alone, due to the $k^2$ scaling of the kSZ reconstruction noise. The chosen $\fnl$ value of 5 is near the $1\sigma$ sensitivity for this configuration. All power spectra are shown at $z=1$ and for mu $|\mu|=1$ (radial modes), and the galaxy power spectra include the RSD term.} 
\label{fig:pggfnl}
\end{figure}

The noise depends on the angle of the mode with respect to the line of sight, i.e. $\mu = \widehat{\k} \cdot \n$, but crucially is independent of the magnitude of $k_L$. The reconstructed velocity field can be related to a reconstruction of the density perturbations $\delta_m$ using linear theory. The reconstruction noise on the density field is thus given by
\be\label{eq:kszmatternoise}
N_{mm}^{\rm rec}(k_L,\mu) = \frac{k_L^2}{(faH)_*^2} N_{vv}^{\rm rec}(k_L,\mu).
\ee
Crucially, the noise is proportional to $k_L^2$, implying that the reconstruction noise on the density field is {\em lowest} on the largest scales. This implies that the density field can be reconstructed at a higher fidelity using kSZ tomography than with direct density measurements from a galaxy survey. This is illustrated for an example experimental configuration in Fig.~\ref{fig:pggfnl}, where we have also shown the added large-scale power on the galaxy power spectrum due to $\fnl$.

We note that on the very largest scales one must account for additional contributions to the kSZ effect beyond the peculiar velocity~\cite{Zhang10d,Deutsch:2017ybc,Terrana2016,Zhang:2015uta}. Taking these into account, kSZ tomography can be used to reconstruct the dipole field, the CMB dipole observed at each point in spacetime. This will modify the noise in Eq.~\ref{eq:kszmatternoise} at very small $k_L$, somewhat increasing the attainable velocity reconstruction noise. This effect will be most important for the deepest galaxy surveys, and we defer further exploration of this point to future work.

\section{Power spectra and experiments}
\label{sec:experiments}

The input data for our forecast are galaxy number densities, biases and the effective beam and noise in an overlapping CMB survey, as well as the small scale power spectra which determine the velocity reconstruction noise. In this section we describe these parameters in detail.

\subsection{Consistent small scale power spectra from the halo model}
  
The kSZ velocity reconstruction noise in eq.\eqref{eq:Nvv} depends on the ratio of small scale power spectra $\left( \frac{P_{ge}(k_S)^2}{P_{gg}^{\rm tot}(k_S) C_l^{\rm tot}} \right)$, where for $\ell>4000$ the CMB power spectrum $C_\ell$ is dominated by the kSZ effect, which depends on $P_{ee}(k_S)$. There is some uncertainty in the shape and amplitude of these three power spectra. However, all three of them can be calculated in the halo model and are dominated by the 1-halo term at the relevant $k_S$. We review the halo model calculation of these power spectra in App.~\ref{app:halomodel}, and more details can be found in~\cite{kszbispectrumpaper}. A key property is that they depend on the satellite galaxy profile in the halo $u_s(k|m,z)$ (assumed NFW, tracing the dark matter) and the electron profile $u_e(k|m,z)$. 
We therefore make a consistent forecast by using a halo model calculation for all three power spectra. 

To calculate galaxy power spectra in the halo model, one needs to specify the Halo Occupation Distribution (HOD). Details about the HOD~\cite{2012ApJ...744..159L,Leauthaud:2011zt} which we use can be found in~\cite{kszbispectrumpaper}. To connect the HOD with different experiments, we are using the following prescription. In the HOD, the galaxy sample is specified by imposing a threshold stellar mass $m_\star^{\rm thresh}$ of observable galaxies. At a fixed halo mass, it assumes a log-normal distribution for the stellar mass. There are also three further parameters in the HOD, which define the central and satellite galaxy numbers for each mass. These parameters depend on $m_\star^{\rm thresh}$, and have been calibrated with data in~\cite{2012ApJ...744..159L}. We match the parameter $m_\star^{\rm thresh}$ so that the total predicted galaxy number (centrals+satellites) matches the number density expected for a given experiment (e.g. LSST, DESI). An example of this matching is shown in the next section.

\subsection{Experiments}
\label{subsec:experiments}

We make forecasts for two next generation large-scale structure experiments, LSST and DESI. LSST is an example of a high number density experiment with photometric redshifts. DESI is an example for a lower number density experiment but with precise spectroscopic redshifts. For the CMB experiment we consider a CMB-S4 configuration~\cite{CMBS42016}, as well as a configuration similar to that of Simons Observatory (SO). We do not include atmospheric noise or noise from foregrounds such as tSZ or CIB in this work. A more realistic forecast that includes these contributions for SO can be found in~\cite{Ade:2018sbj}. Our detailed redshift binned forecast will be for LSST+CMB S4, which is the most promising configuration for $\fnl$, while for DESI+SO we only provide a simplified forecast to illustrate the performance of a lower number density without photo-z errors.

\subsubsection{Large scale structure experiments}

\begin{figure}[tbh]
  \includegraphics[width=0.45\textwidth]{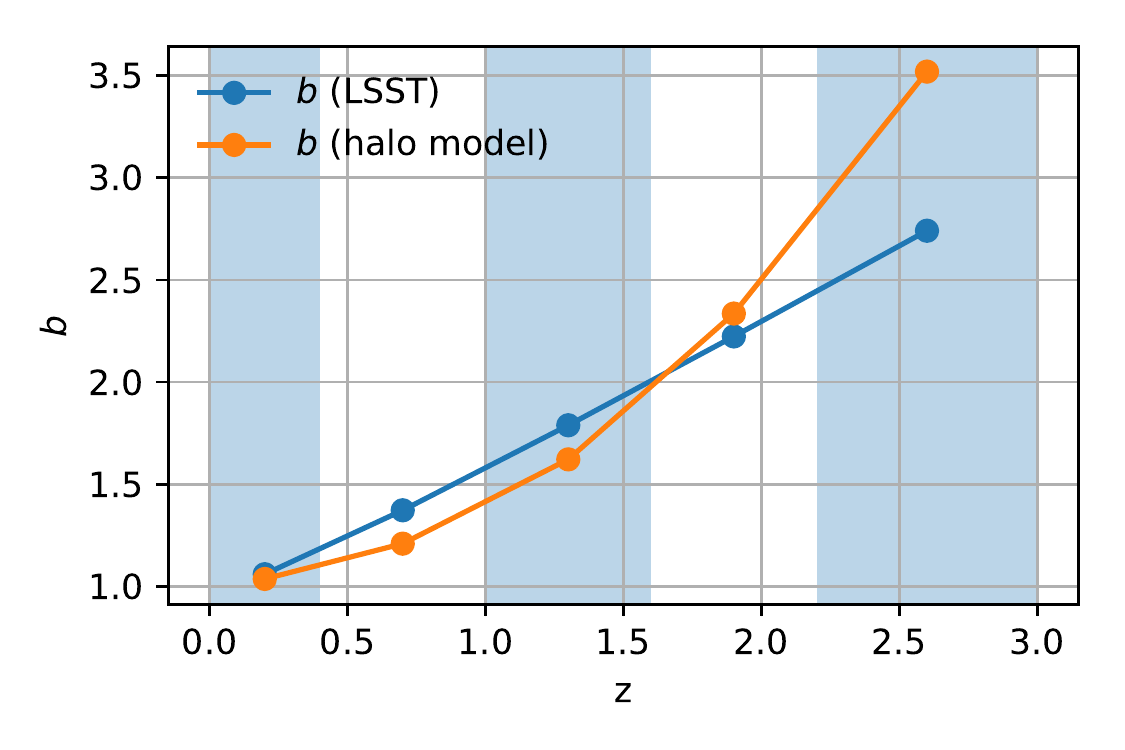}\includegraphics[width=0.45\textwidth]{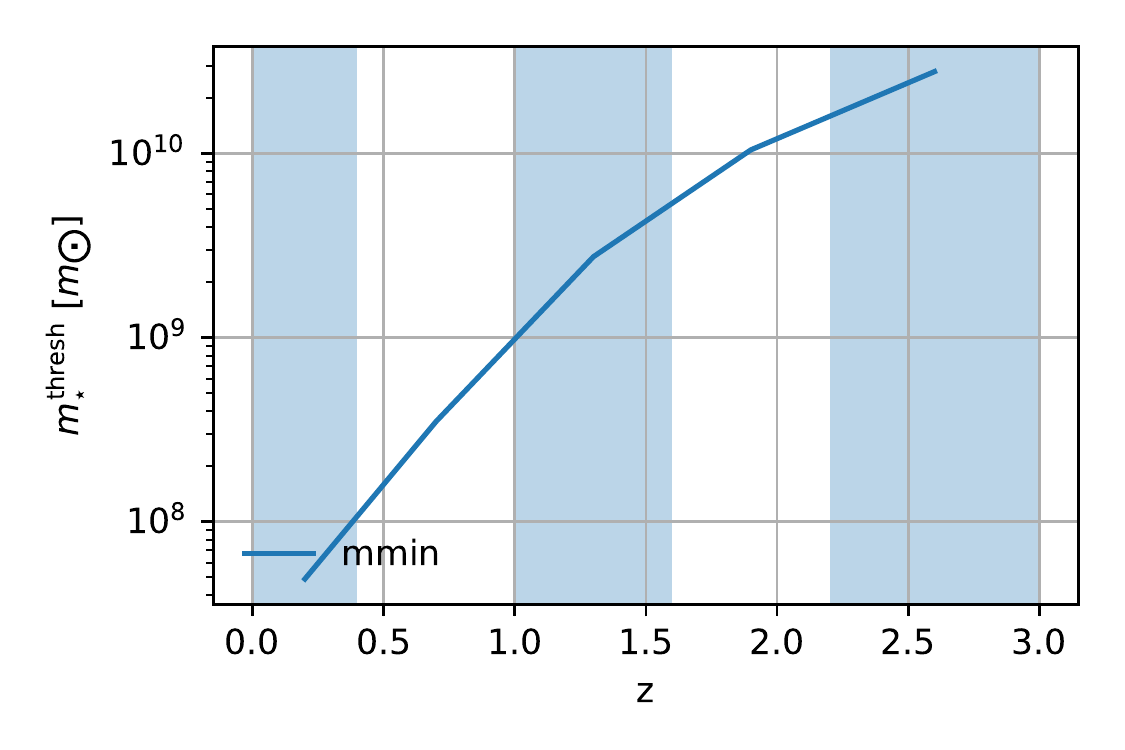}
  \caption{Left: redshift binning and bias as a function of $z$ for the LSST forecast. Right: HOD stellar mass threshold $m_\star^{\rm thresh}$ as a function of $z$ matched so that the galaxy number predicted by the halo model matches Eq.~\eqref{eq:lsstnr}, extracted from the LSST science book.} 
\label{fig:lsstparams}
\end{figure}

Our forecast for Large Synoptic Survey Telescope (LSST) is based on the \textit{LSST Gold Sample} as defined in the LSST science book~\cite{2009arXiv0912.0201L}, which is used in the clustering forecasts. For this dataset, the galaxy number density $n$ per arcmin$^2$ is described by
\be
n(z) =  n_{\rm gal} \ \frac{1}{2 z_0} \left(\frac{z}{z_0}\right)^2 \exp(-z/z_0)  \label{eq:lsstnr}
\ee
with $z_0 = 0.3$ and $n_{\textrm{gal}}=40 \ {\rm arcmin}^{-2}$. The predicted photo-z error is
\be
\sigma_z = 0.03 \ (1+z)
\ee
For the same sample the LSST group also provides the bias
\be
b(z) = 0.95/D(z)
\ee
with the growth factor normalized as $D(z=0) = 1$. The bias with this prescription is plotted in Fig.~\ref{fig:lsstparams}, for five large redshift bins defined below. For comparison, we also show the halo model bias prescription, obtained by adjusting the mass threshold of the HOD to match the number densities provided by LSST. The agreement between the two methods of bias determination is not perfect but within the spread of what is expected for the uncertainty of the bias of a galaxy sample. Below we use the LSST bias in our LSST forecast, to facilitate comparison with other studies. We use the matched mass threshold of the HOD to compute the small scale power spectra $P_{gg}$ and $P_{ge}$ that appear in the velocity reconstruction noise. It should be noted that the bias is a significant uncertainty in the $\fnl$ forecast that can influence results up to a factor of $2$. This uncertainty dominates over the uncertainty in $N_{vv}$ from different small scale power spectra, because the $\fnl$ forecast is dominated by the galaxy shot noise.

For the Dark Energy Spectroscopic Instrument (DESI)~\cite{Aghamousa:2016zmz}, we make a simplified forecast ignoring redshift evolution. Our assumption will be a galaxy density $n_{\textrm{gal}} = 10^{-4} \mathrm{Mpc}^{-3}$ with a bias $b=1.6$ at central redshift $z_\star = 1$, roughly in line with the DESI whitepaper~\cite{Aghamousa:2016zmz}, with perfect redshifts.

\subsubsection{CMB experiments}

Our baseline CMB survey is the planned CMB-S4 experiment. While a definitive instrument specification is still pending, we consider one of many possible configurations that result in a foreground-cleaned CMB map with an effective beam full-width-half-maximum (FWHM) of 1.5 arcminutes and an effective white noise level of 1.0 $\mu K$-arcmin. We do not include atmospheric $1/f$ noise as it is expected to be subdominant to instrument and kSZ contributions at the relevant high multipoles of $\ell>4000$. The total CMB noise that enters Eq. \ref{eq:Nvv} is then

\be
C_l^{\rm tot} = C_l^{\mathrm{TT}} + C_l^{\mathrm{kSZ-reionization}} + C_l^{\mathrm{kSZ-late-time}} + N_l
\ee

where $C_l^{\mathrm{TT}}$ is the lensed CMB temperature power spectrum, $C_l^{\mathrm{kSZ-reionization}}$ is the reionization contribution to kSZ and $C_l^{\mathrm{kSZ-late-time}}$ is the late-time (low-redshift) contribution to kSZ and $N_l$ is the beam-deconvolved noise spectrum of the foreground-cleaned CMB map.

\be
N(\ell) = s_{w}^2\rmn{exp}\left(\frac{\ell(\ell+1)\theta_{\rmn{FWHM}}^2}{8\rmn{ln}2}\right).
\ee

While we focus on CMB-S4, we will also forecast a configuration with noise and beam comparable to Simons Observatory, with a beam FWHM of 1.5 arcminutes and an effective white noise level of 5.0 $\mu K$-arcmin.

\section{Fisher forecast setup}
\label{sec:fishersetup}

We now describe our Fisher forecast setup, including the relevant systematics. We first discuss the simpler case without mass binning of galaxies or halos, then make some analytic approximations, and finally generalize to a mass binned tracer.

\subsection{Methodology}

Our measured data are the modes $(v_\k, \delta^h_\k)$, where the $v_\k$ are the modes coming from the kSZ velocity field reconstruction and $\delta^h_\k$ are the modes of the galaxy survey (assuming a single galaxy per halo). The signal and noise covariance matrices are thus
\begin{align}
\mbf{S}(k,\mu,z) = 
  \begin{pmatrix}
    P_{vv} & P_{vh}\\
    P_{vh} & P_{hh}
  \end{pmatrix},
\hspace{2cm}
\mbf{N}(k,\mu,z) = 
  \begin{pmatrix}
    N_{vv} & 0 \\
    0 & N_{hh}
  \end{pmatrix},
\end{align}
where the noise covariance matrix is diagonal. The total covariance is
\be
\label{eq:covmat1}
\mbf{C}(k,\mu,z) = \mbf{S}(k,\mu,z) + \mbf{N}(k,\mu,z)
\ee
The Fisher matrix for a single redshift bin at $z_*$ is 
\ba
F_{ab} &=& \frac{V}{2} \int \frac{d^3k}{(2 \pi)^3} \text{Tr} \left[ \mbf{C}(\k)_{,a} \mbf{C}(\k)^{-1} \mbf{C}(\k)_{,b} \mbf{C}(\k)^{-1}\right] \\
&=& \frac{V}{2} \int_{k_{\rm min}}^{k_{\rm max}} \int_{-1}^{1} \frac{2\pi \ k^2 \ dk \ d\mu}{(2 \pi)^3} \text{Tr} \left[ \mbf{C}(k,\mu)_{,a} \mbf{C}(k,\mu)^{-1} \mbf{C}(k,\mu)_{,b} \mbf{C}(k,\mu)^{-1}\right], 
\ea
where we have taken into account the $\mu$ angle dependence induced by the kSZ reconstruction. We also take into account redshift space distortions and a free normalization parameter of the velocity reconstruction $b_v$, which is associated with the kSZ optical depth degeneracy\footnote{The kSZ optical depth degeneracy is the fact that the overall normalization of the electron profile in a halo is not known very well, and leads to an unknown overall normalization of the measured velocity field. Mathematically, in the kSZ bispectrum formalism~\cite{kszbispectrumpaper}, a constant factor can be moved between $P_{ge}$ and $P_{gv}$ in the squeezed limit bispectrum without changing its shape.}~\cite{kszbispectrumpaper}. The relevant power spectra become
\ba
P_{hh}(k, z, \mu) &=& \left( b_h +\fnl \frac{\beta_f}{\alpha(k,z)} + f \mu^2 \right)^2 P_{mm}(k,z)  \label{eq:powerspectra1}\\
P_{vh}(k, z, \mu) &=& \left( \frac{b_v f a H}{k} \right) \left( b_h + \fnl \frac{\beta_f}{\alpha(k,z)} + f \mu^2  \right)  P_{mm}(k,z) \\
P_{vv}(k, z) &=& \left( \frac{b_v f a H}{k} \right)^2 P_{mm}(k,z).
\ea
The noise power spectra are given for the velocity reconstruction in Eq.~\eqref{eq:Nvv} and for halos by the shot noise $N_{hh} =  \frac{1}{n_{h}}$ with halo density $n_{h}$. 

We also consider photo-z errors. These can be implemented for halos by a convolution of the halo density field with a Gaussian kernel in radial direction. The halo noise power is then
\be
N_{hh}(k,\mu) = \frac{1}{W^{2}(k,\mu) \ n_{h}} 
\ee
where 
\be
W ^2(k,\mu) = e^{-k^2 \mu^2 \sigma^2(z) / H^2(z)}
\ee
with redshift scattering $\sigma(z)$. The noise in the kSZ velocity reconstruction due to photo-z errors is discussed in detail in~\cite{kszbispectrumpaper} and is also included in this forecast. We further assume that the Volume $V$ limits the available largest modes in the Fisher forecast as $k_{\rm min} = \frac{\pi}{V^{1/3}}$. In our $\fnl$ forecast, we marginalize over $b_h$ and $b_v$. We have also experimented with marginalizing over cosmological parameters, but found that these do not significantly change the sensitivity, as the $\fnl$ distortion of the power spectrum is orthogonal to changes induced by cosmological parameters.

\subsection{Analytic approximation of the Fisher matrix}

To gain some analytic insight into the expected behavior of the signal to noise, we analyze the diagonal $\fnl$ term of the Fisher matrix, based on the analysis for the lensing-galaxy cross-correlation in~\cite{Schmittfull:2017ffw}. To simplify the notation we work with modes $\delta^{\rm rec}_m(\k) = \frac{k}{f a H} v(\k)$, and drop the Kaiser redshift distortion term. The Fisher matrix from a single $\k$ mode is 
\be
  F_{\fnl\fnl}(k,\mu) = \frac{1}{2}\sum_{abcd\in\{m,g\}}
C^{ab}_{,\fnl}(k,\mu)
(C^{-1})^{bc}(k,\mu)
C^{cd}_{,\fnl}(k,\mu)
(C^{-1})^{da}(k,\mu).
\ee
Inserting the covariance matrix Eq.~\eqref{eq:covmat1}, one finds~\cite{Schmittfull:2017ffw} that
\begin{align}
  F_{\fnl\fnl}(k,\mu) =\, & \frac{1}{2\left(1-r^2\right)^2}\Bigg[ 
\left(\frac{C^{hh}_{,\fnl}}{C^{hh}}-2r^2\frac{C^{mh}_{,\fnl}}{C^{mh}}\right)^2+2r^2 (1-r^2)\left(\frac{C^{mh}_{,\fnl}}{C^{mh}}\right)^2
\,\Bigg].
\end{align}
where we defined the correlation coefficient 
\be
r(k,\mu) = \frac{C_{mh}(k)}{\sqrt{C_{mm}(k,\mu)C_{hh}(k)}}.
\ee
From Eq.~\eqref{eq:powerspectra1}, for fiducial $\fnl=0$, we have
\begin{align}
\frac{C^{hh}_{,\fnl}}{C^{hh}}(k) &= \frac{2 b_h \frac{\beta_f}{\alpha} P_{mm}}{b_h^2 P_{mm} + N_{hh}} \nn \\
\frac{C^{mh}_{,\fnl}}{C^{mh}}(k) &= \frac{\beta_f}{b_h \alpha}
\end{align}
In the limit $P_{hh} \gg N_{hh}$ we find
\be
F_{\fnl\fnl}(k,\mu) =  \frac{2-r^2}{1-r^2} \left(\frac{\beta_f}{b_h \alpha}\right)^2
\ee
which for $r\rightarrow1$ scales as $(1-r^2)^{-1}$ (as found for sample variance cancellation in~\cite{Seljak:2008xr}), leading to a large decrease in uncertainty on $\fnl$ as $r$ approaches 1. Knowing $r(k,\mu)$ we can estimate the Fisher matrix as
\be
F_{\fnl\fnl} = \frac{V}{2} \int_{k_{\rm min}}^{k_{\rm max}} \int_{-1}^{1} \frac{k^2 \ dk \ d\mu}{(2 \pi)^2} F_{\fnl\fnl}(k,\mu).
\ee
In the case of a configuration with strong sample variance cancellation (i.e. almost all information comes from the cross-correlation), this approximation is almost exact. This is the case for the baseline 2 experiment to be defined below, while for the baseline 1 experiment with less sample variance cancellation the estimate is about $15\%$ too large compared to the full answer given below. 

To understand the behavior of the cross correlation coefficient better, we can write it for $b_h \sim 1$ as
\be
\label{eq:rfactor}
r \sim \Big[ \Big( 1+\frac{N^{\rm rec}_{mm}}{P_{mm}} \Big) \Big(1+\frac{N_{hh}}{P_{hh}} \Big) \Big]^{-{1/2}}  
\ee
We see that even in the limit $\frac{N^{\rm rec}_{mm}}{P_{mm}} \rightarrow 0$, the best our method could possibly achieve, the correlation coefficient and thus the sensitivity to $\fnl$ is limited by the halo shot noise. We will plot the correlation coefficient below for different experimental configurations, finding very encouraging results. The importance of the correlation coefficient for sample variance cancellation strongly suggests the power of our method for $\fnl$ determination.

\subsection{Mass binned forecast (multi-tracers)}

Where observationally feasible, sample variance cancellation can also be achieved by mass binning galaxies (or more precisely their host halos). This is because the halo bias is a function of halo mass, and therefore by measuring the same $\k$-modes with different masses/biases one can again cancel the stochastic mode amplitude. To explore the influence of this effect, we provide a forecast assuming that this highly non-trivial procedure can be done perfectly. The measured data is now the set of modes $(v_{\k,1}, ... , v_{\k,N}, \delta^h_{\k,1} , ... , \delta^h_{\k,N})$, i.e. we measure a kSZ velocity reconstruction and a halo distribution in each mass bin. In each bin $i$ the mean (number weighted) bias is given by
\be
b_{h,i} = \frac{\int_{M \in {\rm bin}\ i} dM \frac{dn}{dM} b_h(M)} {\int_{M \in {\rm bin}\ i} dM \frac{dn}{dM}}
\ee
In addition each mass bin has its own free velocity normalization $b_{v,i}$, which corresponds to the kSZ optical depth degeneracy discussed above. The signal covariance matrix is now
\be
\mbf{S}(k,\mu,z) = 
  \begin{pmatrix}
    \mathbf{P}_{vv} & \mathbf{P}_{vh}\\
    \mathbf{P}_{vh} & \mathbf{P}_{hh}
  \end{pmatrix}
\ee
with
\ba
P_{hh,ij}(k, z, \mu) &=& \left( b_{h,i} +\fnl \frac{\beta_f}{\alpha(k,z)} + f \mu^2 \right) \left( b_{h,j} +\fnl \frac{\beta_f}{\alpha(k,z)} + f \mu^2 \right) P_{mm}(k,z)  \label{eq:powerspectra2} \\
P_{vh,ij}(k, z, \mu) &=& \left( \frac{b_{v,i} f a H}{k} \right) \left( b_{h,j} + \fnl \frac{\beta_f}{\alpha(k,z)} + f \mu^2  \right)  P_{mm}(k,z) \\
P_{vv,ij}(k, z) &=& \left( \frac{b_{v,i} f a H}{k} \right) \left( \frac{b_{v,j} f a H}{k} \right) P_{mm}(k,z).
\ea
The noise power is
\be
\mbf{N}(k,\mu,z) = 
  \begin{pmatrix}
    \mathbf{N}_{vv} & 0 \\
    0 & \mathbf{N}_{hh}
  \end{pmatrix}
\ee
Here $\mathbf{N}_{hh}$ is given by the halo shot noise
\be
N_{hh,ij} = \frac{1}{W^{2}} \frac{\delta_{ij}}{{n_{h,i}}} \label{eq:Eij}.
\ee
This Poisson term is the dominant halo noise term and the only one we considered here (see~\cite{Ferraro:2014jba} for a discussion of corrections including off diagonal noise between mass bins). The velocity reconstruction noise $N_{vv,ij}(k,\mu)$ is given as follows. First we define:
\be
A_{ij} = \int dk_S \, k_S \left( \frac{P_{ge,i}(k_S) P_{ge,j}(k_S) P_{gg,ij}^{\rm tot}(k_S)}{P_{gg,ii}^{\rm tot}(k_S) P_{gg,jj}^{\rm tot}(k_S) C_l^{\rm tot}} \right)_{l=k_S\chi_*}
\ee
where we have introduced the notation $P^{\rm tot}_{gg,ij} = P_{gg,ij} + N_{gg,ij}$ for the total (clustering + Poisson)
galaxy power spectrum, and $P_{ge,i}$ for the electron-galaxy cross spectrum.
Then:
\be
N_{vv,ij}(k,\mu) = \mu^{-2} \frac{2\pi \chi_*^2}{K_*^2} \frac{A_{ij}}{A_{ii} A_{jj}}
\ee
As a sanity check, for an auto power spectrum $(i=j)$, the result is the same as before in Eq.~\eqref{eq:Nvv}.

\section{Fisher forecast results} 
\label{sec:fisherresults}

In this section we provide Fisher forecasts for different experimental setups. In the first part, we analyze two realistic baseline configurations for a single redshift bin and without mass binning in detail. We then add redshift binning, to obtain a realistic forecast for LSST. Finally we investigate the influence of mass binning halos, where sample variance cancellation already appears at the level of galaxies alone, and the improvement factor is thus reduced.

\subsection{Baseline forecast: single 3d snapshot box, no mass binning}

To explore the parameter dependencies of our forecast, we start with the two baseline experiments specified in Table~\ref{tab:baseline}. These baseline values were chosen to resemble the experimental configuration of DESI and an SO-like CMB experiment (baseline 1) and of LSST and CMB-S4 (baseline 2). For simplicity here we have used a single 3-dimensional box in our kSZ box formalism, where the box has the size of the survey volume. Therefore the forecast in this section ignores the time evolution of power spectra and biases on the light cone, but retains the unbinned red shift (or distance) information of the galaxies. In the next section, we approximate light cone evolution by using a sequence of boxes of the appropriate volume for a series of redshift bins along the light cone. A precise treatment of light cone evolution would require using spherical coordinates and is postponed to future work.

For the baseline 1 experiments, we forecast a combined constraint $\sigma_{f_{NL}}^{\rm kSZ+gal} = 3.3$, an improvement factor of $1.8$ with respect to the galaxy value $\sigma_{f_{NL}}^{\rm gal} = 6.0$. For the baseline 2 experiments we find $\sigma_{f_{NL}}^{\rm kSZ+gal} = 0.7$ and $\sigma_{f_{NL}}^{\rm gal} = 5.3$ with an improvement factor of $7.8$. Note that this large improvement factor is reduced when considering all redshifts or considering mass binning below. The forecasts shows that the kSZ method benefits strongly from a high number density, and is not very sensitive to photo-z errors. 

\begin{table}[h!]
\centering
\begin{tabular}{ccc}
\ \ & baseline 1 & baseline 2 \\
\hline
survey volume $V$\ \ \ \ & $100 \ \mathrm{Gpc}^3$ & $100 \ \mathrm{Gpc}^3$ \\ 
central redshift $z$\ \ \ \ & $1.0$ & $1.0$ \\ 
galaxy density $n_g$\ \ \ \ & $2 \times 10^{-4} \ \mathrm{Mpc}^{-3}$ & $10^{-2} \ \mathrm{Mpc}^{-3}$ \\ 
halo bias $b_h$ \ \ \ \ & $1.6$ & $1.6$  \\ 
photo-z error $\sigma_z$ \ \ \ \ & - & 0.06\\ 
CMB sensitivity\ \ \ \ & $5\mu$K-arcmin & $1\mu$K-arcmin \\
CMB resolution\ \ \ \ & 1.5' & 1.5' \\ 
$\sigma_{f_{NL}}^{\rm gal}$ \ \ \ \ & 6.0 & 5.3 \\
$\sigma_{f_{NL}}^{\rm kSZ+gal}$\ \ \ \ & 3.3 & 0.7 \\
$\sigma_{f_{NL}}^{\rm gal}/\sigma_{f_{NL}}^{\rm kSZ+gal}$\ \ \ \ & 1.8 & 7.8\\
\hline
\end{tabular}
\caption{Baseline configuration of LSS and CMB experiments. The values for baseline 1 are similar to those expected for DESI and Simons Observatory. The values for baseline 2 are similar to LSST and CMB S4. Bias and survey volume were kept identical for both baselines to stress the dependence on galaxy density and photo-z errors. Here we only consider one 3-dimensional redshift box. For the full redshift range of LSST, including the whole survey volume with sky overlap with CMB S4, see below.}
\label{tab:baseline}
\end{table}

\begin{figure}[tbh]
  \includegraphics[width=0.45\textwidth]{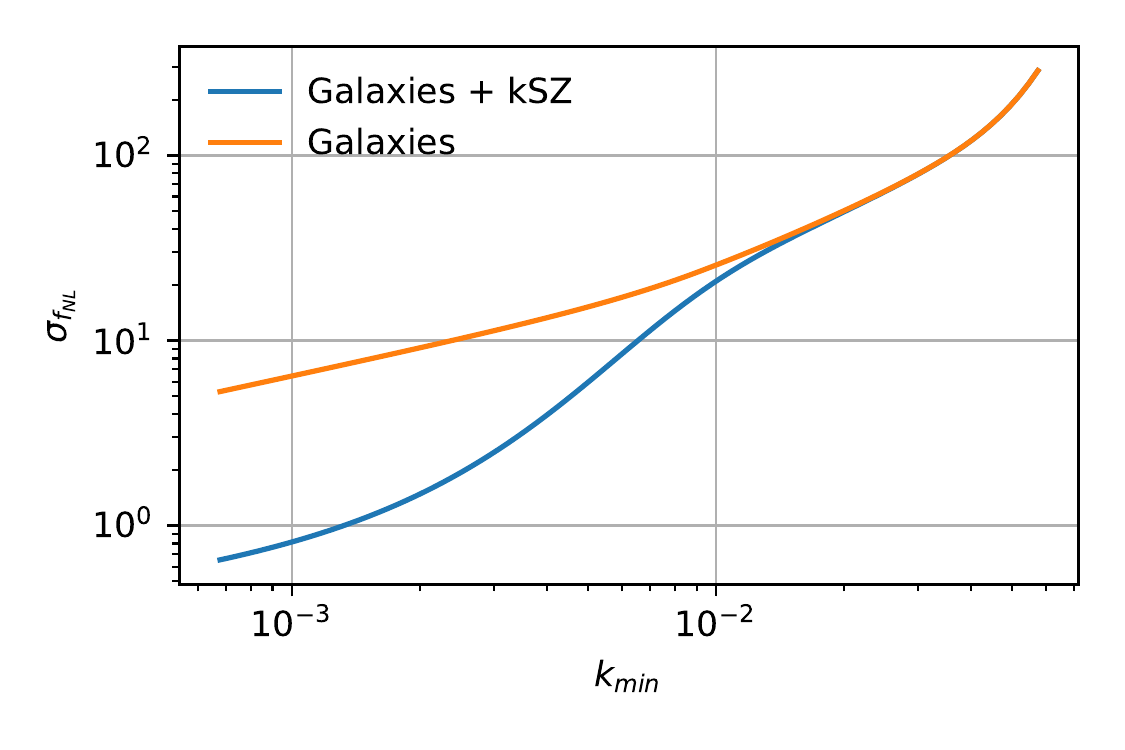} \includegraphics[width=0.45\textwidth]{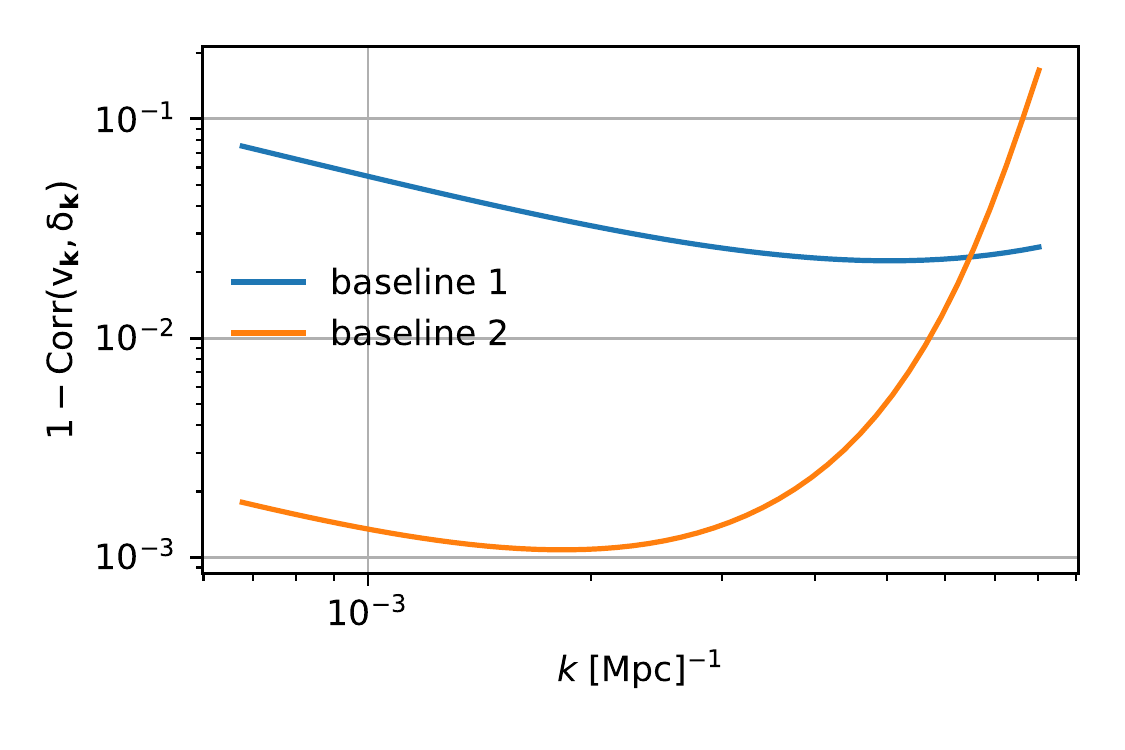}
  \caption{Left: Sensitivity $\sigma_{f_{NL}}$ as a function of $k_{\rm min}$ for baseline 2. On the lower end, $k$ is cut off by the size of the volume. The plot illustrates which scales contribute most to the signal. Right: Correlation coefficient of velocity and galaxy modes for $\mu=1$ modes, for both baseline configurations. The influence of photo-z errors in baseline 2 is clearly visible at high $k$.} 
\label{fig:kminscaling}
\end{figure}

\begin{figure}[tbh]
  \includegraphics[width=0.45\textwidth]{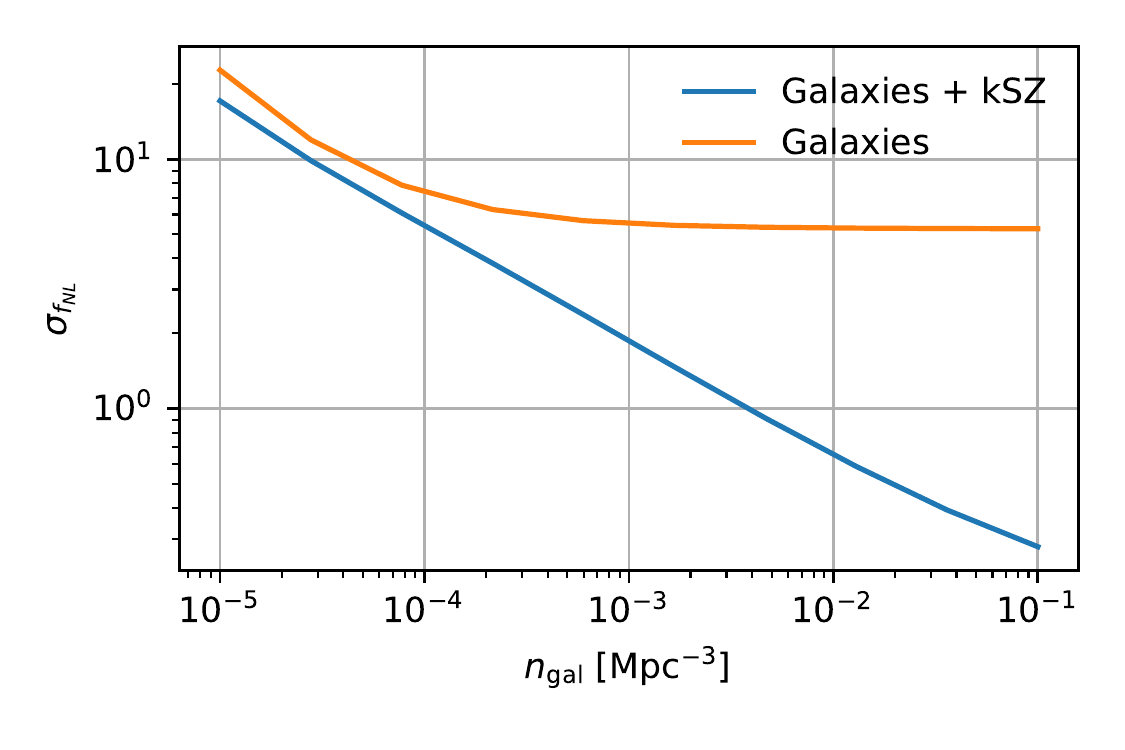} \includegraphics[width=0.45\textwidth]{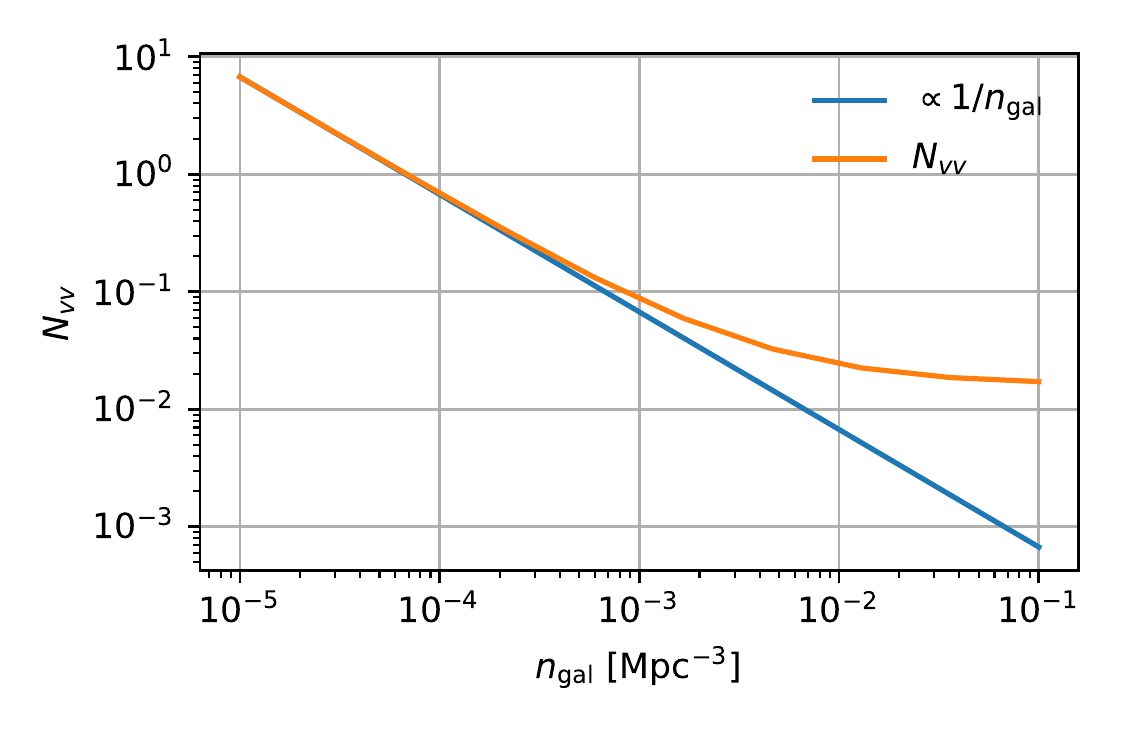} 
  \caption{Left: Sensitivity $\sigma_{f_{NL}}$ as a function of galaxy density $n_{\rm gal}$ for baseline 2. Galaxy density is a critical parameter for our method, in particular because increasing it improves the shot noise of the galaxy mode for sample variance cancellation. For clarity in this plot we have kept the bias constant, however of course highly biased galaxies are limited in number. Right: Velocity reconstruction noise $N_{vv}$ as a function of galaxy density $n_{\rm gal}$ for baseline 2. This illustrates that for baseline 2 the galaxy density is so large that with CMB S4 noise levels we are close to signal-to-noise saturation of the velocity reconstruction as a function of $n_{\rm gal}$.} 
\label{fig:ngalscaling}
\end{figure}

\begin{figure}[tbh]
  \includegraphics[width=0.45\textwidth]{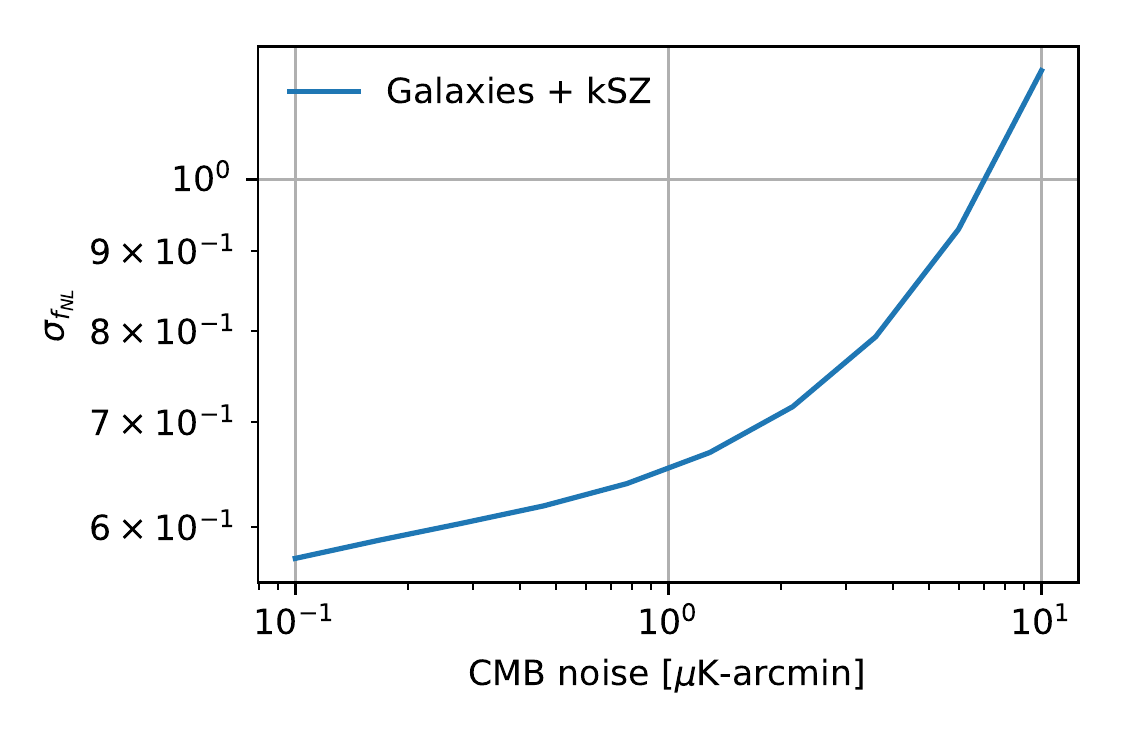} \includegraphics[width=0.45\textwidth]{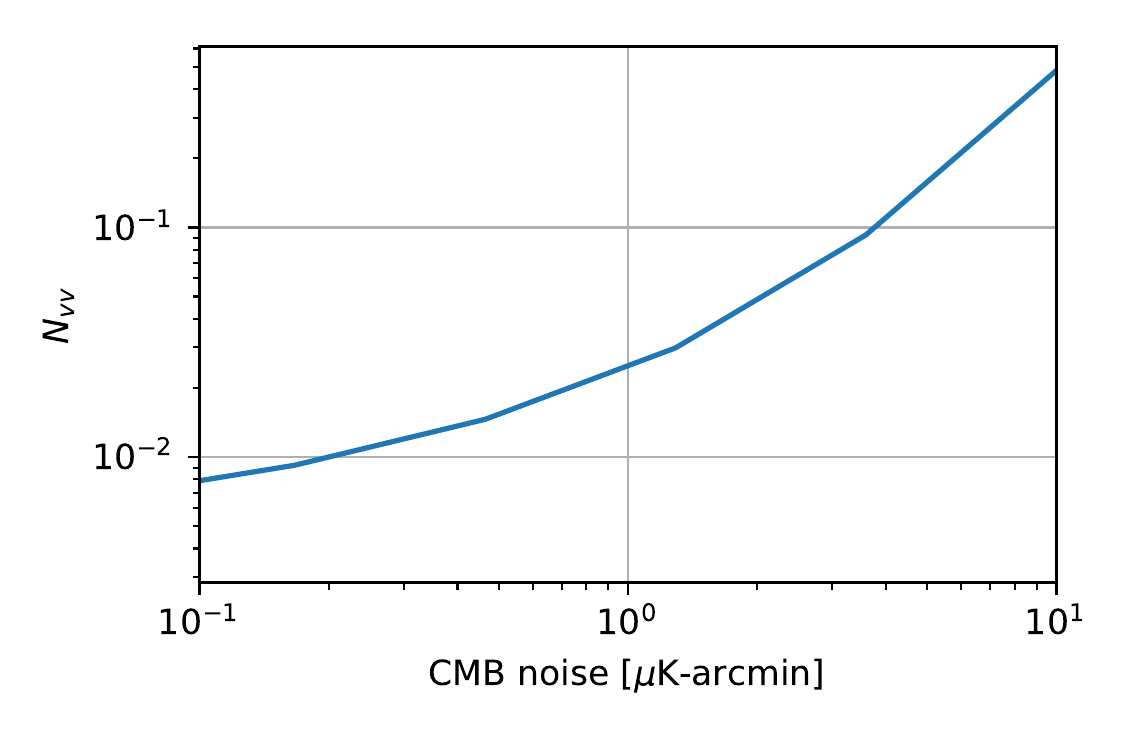} 
  \includegraphics[width=0.45\textwidth]{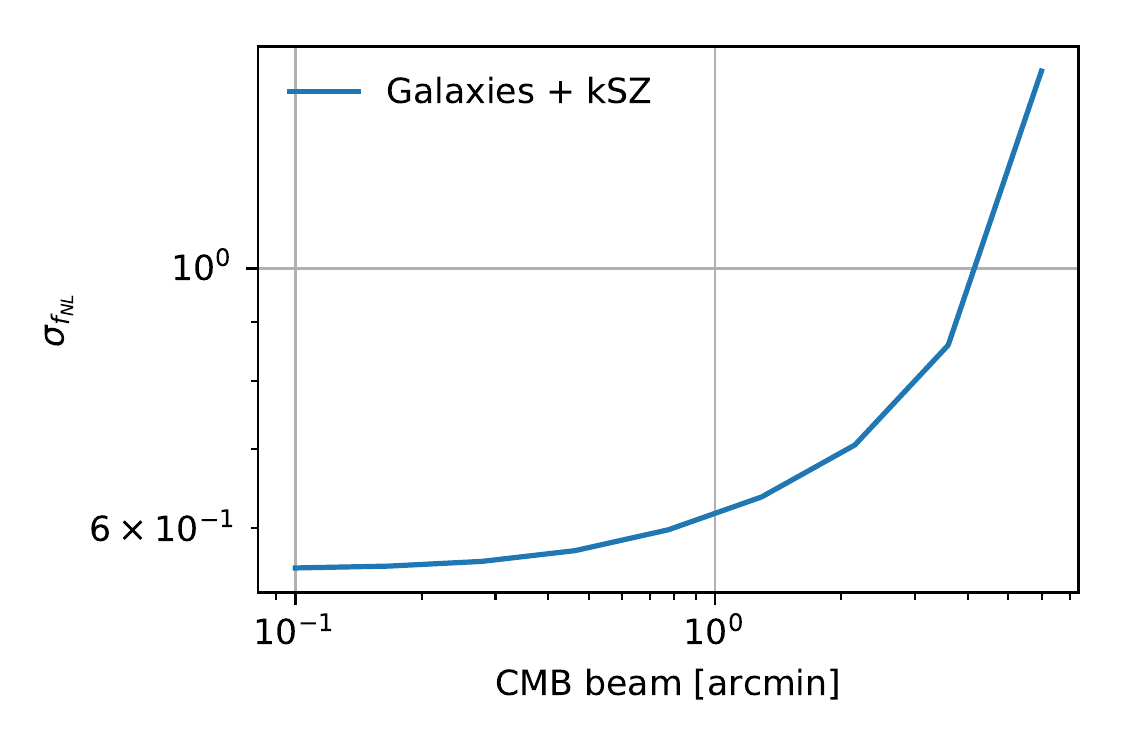} \includegraphics[width=0.45\textwidth]{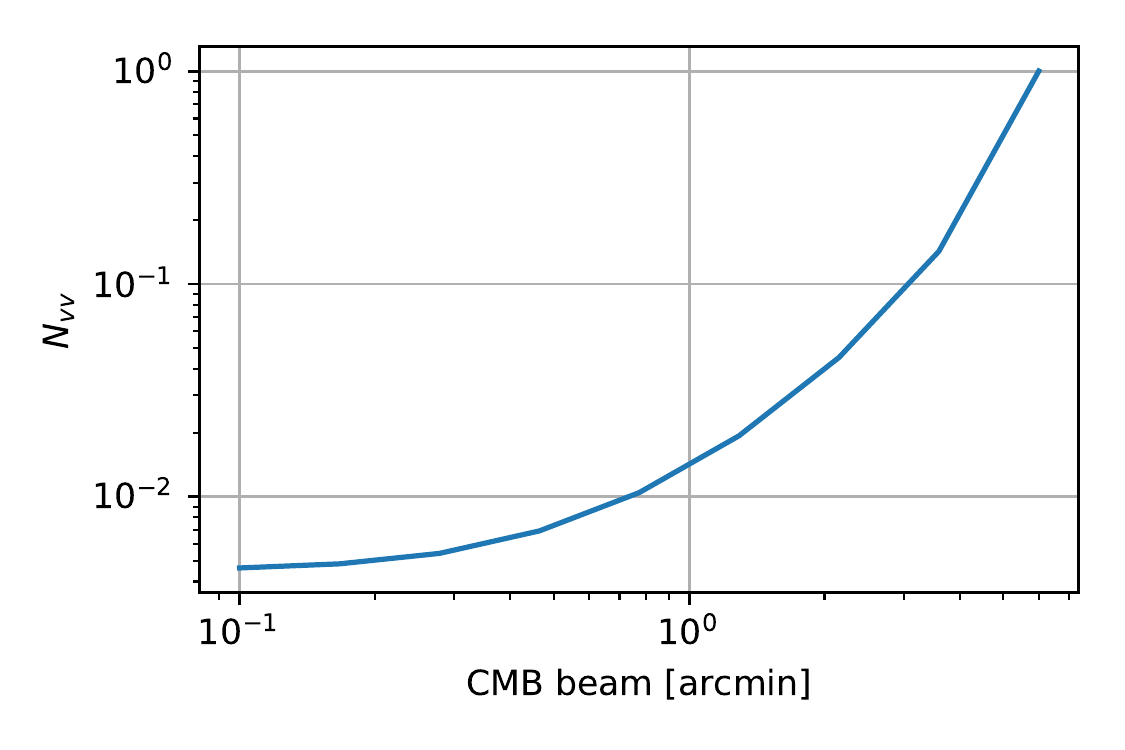} 
  \caption{Top: Sensitivity $\sigma_{f_{NL}}$ (left) and velocity reconstruction noise $N_{vv}$ (right) as a function of CMB noise for baseline 2. Bottom:  Sensitivity $\sigma_{f_{NL}}$ (left) and velocity reconstruction noise $N_{vv}$ (right) as a function of CMB beam for baseline 2. For the $\fnl$ measurement the CMB noise parameters are less critical than the galaxy density, as long as they suffice to get a velocity reconstruction from the CMB, in which case the velocity reconstruction becomes quickly superior to the shot noise limited galaxy mode measurement.} 
\label{fig:cmbscaling}
\end{figure}

To explore which scales contribute most to the signal, we plot $\sigma_{f_{NL}}$ as a function of $k_{\rm min}$ in Fig.~\ref{fig:kminscaling} (left). The plot shows where the effect of kSZ sample variance cancellation kicks in, around $k=0.01 \ \mathrm{Mpc}^{-1}$. It also shows that towards the low $k_{\rm  min}$ end both curves scale similarly with $k_{\rm  min}$ in this configuration. We also plot the correlation coefficient for radial modes in Fig.~\ref{fig:kminscaling} (right). As we have seen, the improvement factor due to sample variance cancellation depends on the correlation coefficient as $(1- \mathrm{Corr}^2)^{-1/2}$. For example, for modes where $\mathrm{Corr} = 0.999$ this gives an improvement factor of $22$ for these modes. This explains the high gain due to kSZ in the baseline 2 configuration.  We further examine the dependence of $\sigma_{\fnl}$ on the galaxy density in Fig.\ref{fig:ngalscaling} and on the CMB experimental data in Fig.~\ref{fig:cmbscaling}. For these plots we have used baseline 2 as our starting configuration, and then varied only the one parameter on the x-axis, keeping all other parameters constant. It is clear that the most critical parameter is the galaxy density. For galaxies alone we quickly enter the cosmic variance limit, so the $\fnl$ signal levels off with respect to galaxy density (Fig.\ref{fig:ngalscaling}). This would be different if we were to mass bin galaxies and get sample variance cancellation from galaxies alone, as we show below in Sec.~\ref{sec:massbinning}. 

We also quantified how much information on $\fnl$ can be obtained without the galaxy auto correlation function. This is important because calibration errors can make the galaxy auto correlation function unreliable on large scales. To implement this, we marginalize over an additional term of form $\gamma/k^2$ in the galaxy auto power, where $\gamma$ is a free parameter, which completely removes any information on $\fnl$ from the galaxy auto power in the forecast. For baseline 1 the forecasted $\sigma_{\fnl}$ increases by only 0.5 percent, and for baseline 2 even by only 0.1 percent. This may seem surprising, especially for baseline 1 where the improvement factor due to sample variance cancellation is moderate, because the galaxy auto power spectrum also includes transverse modes which are not measurable with kSZ tomography. However, due to the extremely low noise in the kSZ velocity field, only a small subset of modes is so transverse that the kSZ velocity noise term $\propto \mu^{-2}$ is near or above the galaxy shot noise. In summary, our method allows one to obtain excellent $\fnl$ constraints without using the galaxy auto correlation function.

\subsection{Redshift evolution for LSST}

We now extend the forecast to cover the entire redshift range of LSST, combined with the CMB S4 mission. Number densities, biases and photo-z errors are as specified in Sec.~\ref{subsec:experiments} for LSST. The bin volumes are chosen so that they include the expected sky overlap of LSST and CMB S4 ($f_{\mathrm{sky}}=0.3$). To take into account redshift dependencies, we have divided the available redshift range $0<z<3$ in five bins, each of which we treat as a 3-dimensional box with the corresponding cosmological volume. The binning parameters are given in Table~\ref{tab:bins}.

\begin{table}[h!]
\centering
\begin{tabular}{cccccc}
bin & $z_{\rm min}$ & $z_{\rm max}$ & halo bias $b_h$ & galaxy density $n_g$ & volume $V$ \\
\hline
1 & 0 & 0.4 & 1.05 & $0.05 \ \mathrm{Mpc}^{-3}$ & $5.2 \ \mathrm{Gpc}^3$ \\     
2 & 0.4 & 1.0 & 1.37& $0.02 \ \mathrm{Mpc}^{-3}$ & $43.6 \ \mathrm{Gpc}^3$ \\ 
3 & 1.0 & 1.6 & 1.79 & $0.006 \ \mathrm{Mpc}^{-3}$ & $75.9 \ \mathrm{Gpc}^3$ \\ 
4 & 1.6 & 2.2 & 2.22 & $0.0015\ \mathrm{Mpc}^{-3}$ & $89.3 \ \mathrm{Gpc}^3$ \\ 
5 & 2.2 & 3.0 & 2.74 & $0.0003 \ \mathrm{Mpc}^{-3}$ & $119.9 \ \mathrm{Gpc}^3$ \\ 
\hline
\end{tabular}
\caption{Redshift binning and survey parameters for the LSST forecast.}
\label{tab:bins}
\end{table}

The Fisher matrix is assumed to be a sum of independent bins of form 
\be
F^{\rm tot} = \sum_i F_{\Delta z_i}
\ee
where $F_{\Delta z_i}$ is the Fisher matrix in each redshift bin and each redshift bin is assumed to have an independent bias $b^i_{h}$ and velocity normalization $b^i_{v}$ to take into account their unknown redshift dependence. One may ask if the approximation of independent redshift bins is sufficient, as the largest scales contribute significantly to the signal as illustrated in Fig.~\ref{fig:kminscaling}. This is indeed the case. Schematically the Fisher matrix is the product of the volume and the $k$-integral, where the volume also limits $k_{\rm min}$. The scaling of the volume  is $V \propto k_{\rm min}^{-3}$ and outruns the scaling $\sim k_{\rm min}^{-1}$ (see Fig.~\ref{fig:kminscaling}) of the $k$-integral in the Fisher matrix on large scales. For this reason, splitting the cosmological volume in a few large independent redshift bins, as we do here, is a reasonable approximation for the total signal-to-noise that can be achieved. 

\begin{figure}[tbh]
  \includegraphics[width=0.45\textwidth]{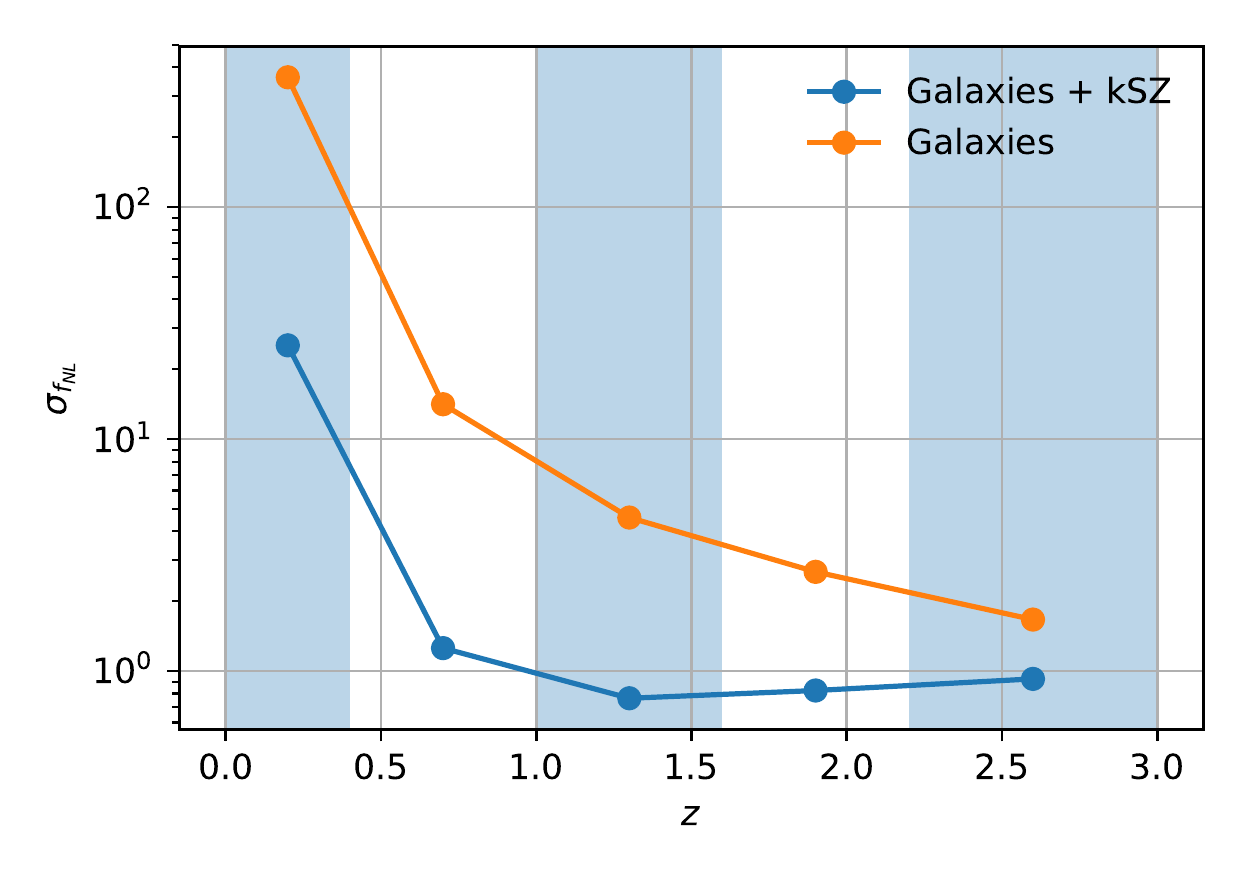}\includegraphics[width=0.45\textwidth]{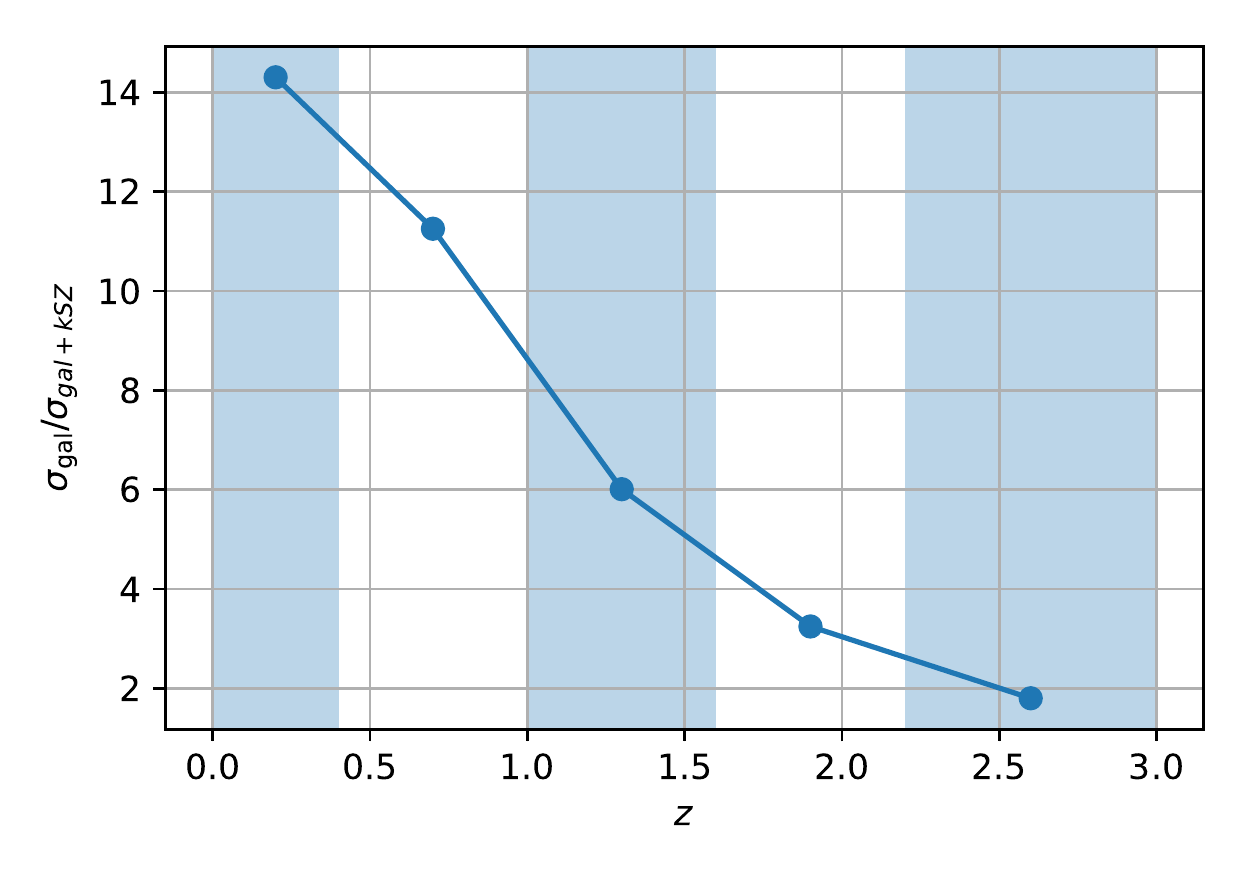}  
  \caption{Constraints $\sigma_{f_{NL}}$ (left) and improvement factor (right) as a function of redshift bin (per bin, not cumulative). The total combined sensitivity for kSZ+galaxies is $\sigma_{f_{NL}}=0.45$, an improvement by a factor of $3.00$ with respect to the galaxies alone with $\sigma_{f_{NL}}=1.35$. The plot includes LSST gold sample galaxy density, biases and photo-z's, and CMB S4 noise levels.} 
\label{fig:fnlforecast2}
\end{figure}

With this approximation we find the results shown in Fig.~\ref{fig:fnlforecast2}. For galaxies, we find that the largest redshift gives the largest signal, which is due to the fact that both the biases and the volumes (for our sampling) grow with $z$, while the falling number densities are not important because we are in the sample variance limited regime. In the case of galaxies+kSZ, there is a competition between the growing biases and volumes and the falling number densities, which are important for sample variance cancellation here. For the total significance of all five bins together we find that for galaxies+kSZ $\sigma_{f_{NL}}=0.45$, an improvement of a factor of $3.0$ with respect to galaxies alone. This results in a potential two sigma exclusion of the multifield limit.

\subsection{Mass binning (multi-tracer forecast)}
\label{sec:massbinning}

As explained above, one can improve $\fnl$ measurements by mass binning galaxies (if accurate masses are available) and thus obtain sample variance cancellation from galaxies alone. Here we forecast combined constraints when using sample variance cancellation both from mass binning and from kSZ. Here we assume abundance matching (one galaxy in each halo) also for small scale power spectra, to avoid mass binning the HOD. This should not change the results qualitatively.

\begin{figure}[tbh]
  \includegraphics[width=0.45\textwidth]{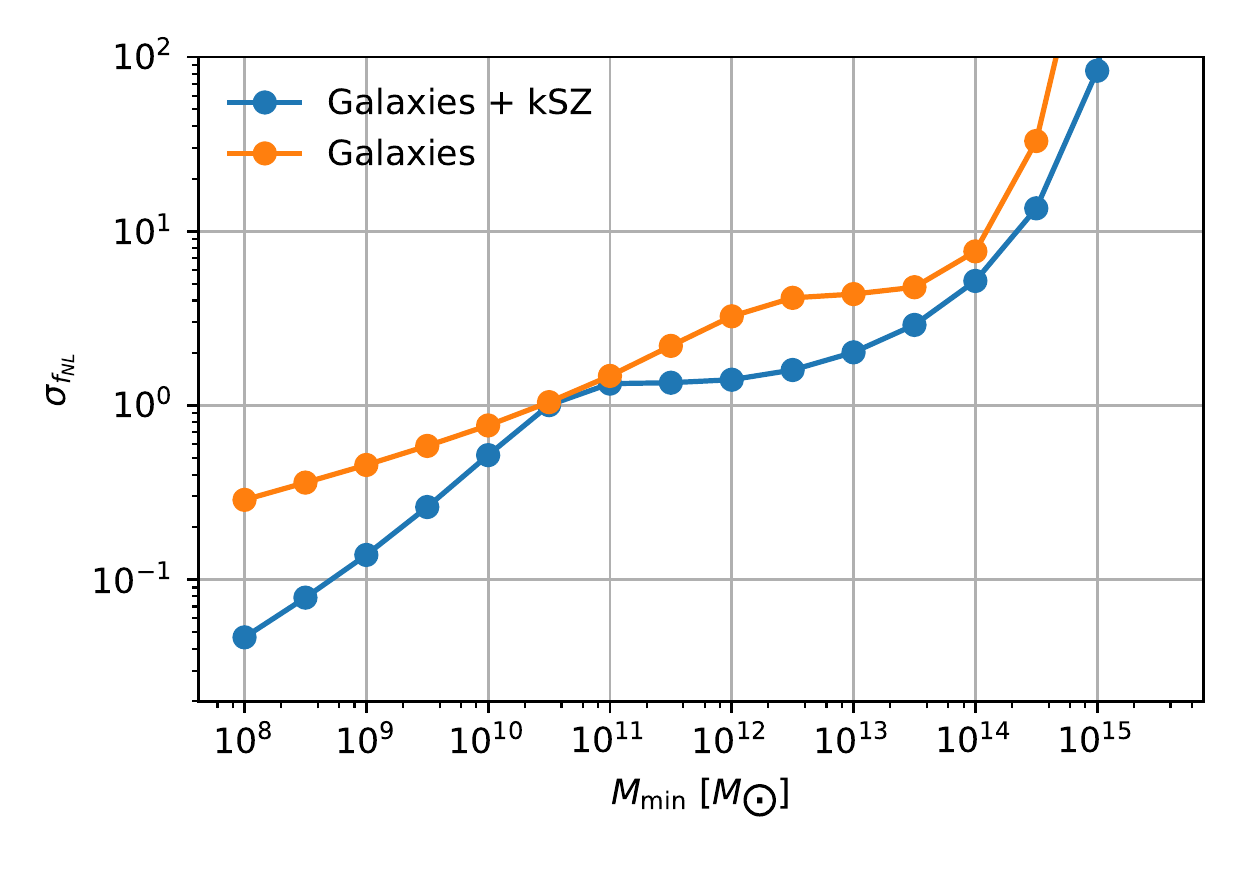}\includegraphics[width=0.45\textwidth]{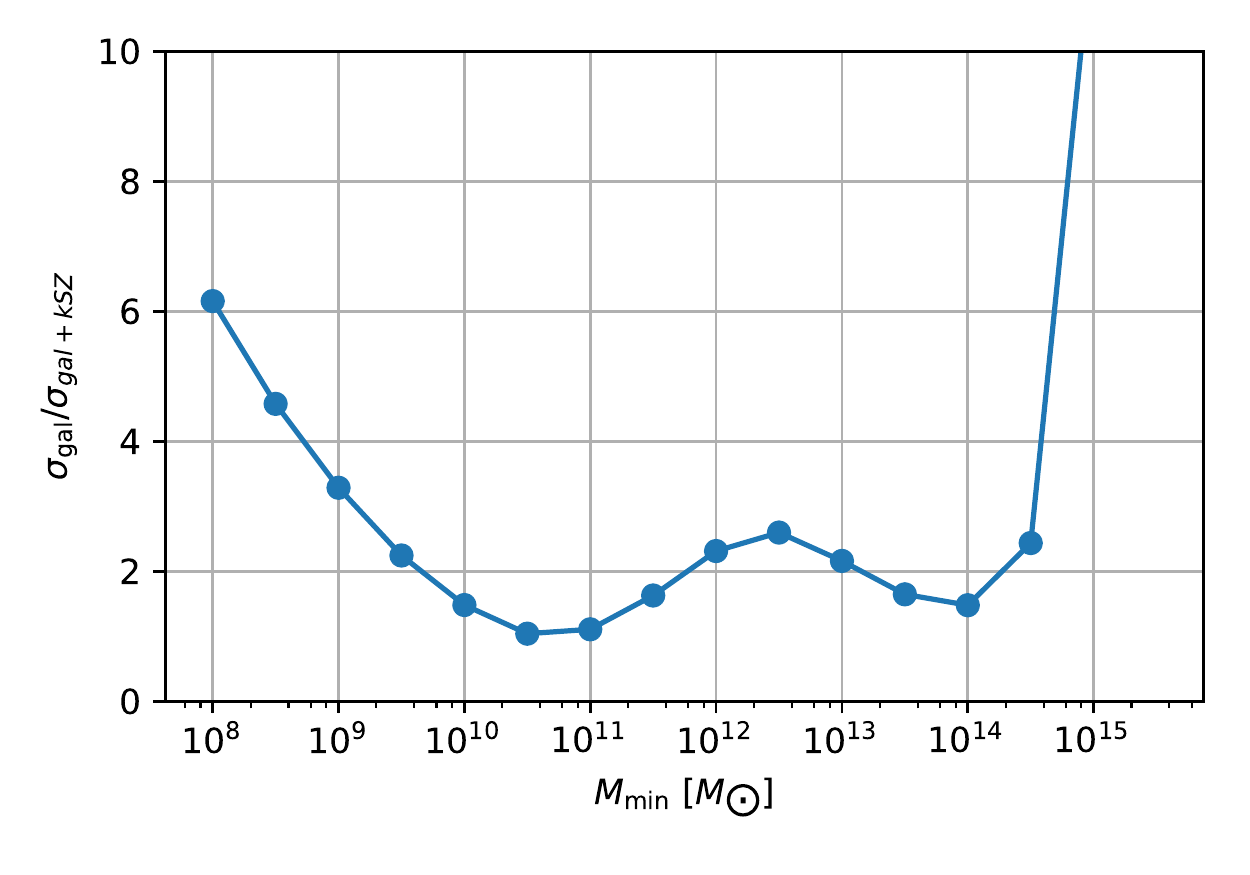}  
  \caption{Left: Constraints $\sigma_{f_{NL}}$ for galaxies and for galaxies+kSZ as a function of $M^{\rm halo}_{\rm min}$ in a multi-tracer analysis. From left to right on the x-axis, each new dot removes one low mass bin from the forecast. For this analysis we used a survey with survey volume $100 \mathrm{Gpc}^3$ and central redshift $z_\star=1$. This forecast uses the halo model number densities $n_h$, biases $b_h$ and power spectra $P_{gg},P_{ge}$. Right: Improvement factor due to adding the kSZ information.} 
\label{fig:fnlforecast3}
\end{figure}

Our $f_{NL}$ forecast results are shown in Fig.\ref{fig:fnlforecast3}, where on the x-axis from the right to the left we continuously add lower mass bins. The mass binning was chosen tight enough so that further binning would not lead to better constraints. For galaxies without kSZ we recover the multi-tracer forecast results of~\cite{Ferraro:2014jba}. In particular the sample variance plateau is visible around $M^{\rm halo}_{\rm min}=10^{13} M_\odot$, where we are limited by cosmic variance but the number densities are not yet large enough for effective sample variance cancellation. For mass cuts in this range, adding kSZ information provides about a factor of two improvement in sensitivity. Interestingly, around $M^{\rm halo}_{\rm min}=10^{11} M_\odot$ the kSZ method provides almost no extra information. This is exactly the halo mass where the halo bias is 1. The reason for the convergence of the two curves around $b=1$ can be understood as follows. The kSZ velocity field provides a very low noise measurement of each mode at $b=1$ for sample variance cancellation. The $\fnl$ information comes from ``comparing'' biased galaxy modes with this $b=1$ reference mode. Galaxy modes that contribute significantly have a bias substantially higher than $1$, with a much larger shot noise than the reference mode. Getting the $b=1$ reference mode from either galaxies or kSZ does not change the signal to noise by much as it is dominated by the higher shot noise of the biased mode (compare the  crucial correlation coefficient in Eq.~\eqref{eq:rfactor}). Another interesting behavior of the mass binned plot is that for halo masses smaller than $M^{\rm halo}_{\rm min}=10^{11} M_\odot$, the kSZ velocity field starts to add information again and scales more favorably than the galaxies alone with respect to $M^{\rm halo}_{\rm min}$. The reason is again that the kSZ provides an almost noise free measurement of the mode at $b=1$, with which the $b<1$ galaxy modes can be compared for sample variance cancellation. Unfortunately this regime will be difficult to exploit in practice, as such halo masses are well below the power of upcoming experiments.

\section{Conclusions}
\label{sec:conclusions}

kSZ tomography with its reconstruction of large scale velocities~\cite{Deutsch:2017ybc,kszbispectrumpaper} is a powerful new probe for cosmology, that will be accessible with the next generation of CMB and large-scale structure experiments. In particular, cross-correlating the velocity field with a galaxy survey leads to sample variance cancellation in the measurement of the galaxy bias and other quantities. In this paper we have worked out in detail how this method can be used to improve constraints on $\fnl$. An application of the same method to constrain additional sources of scale dependence in the galaxy bias and growth rate from massive neutrinos, dark energy perturbations or modified gravity is left for future work.

The statistical power of our method arises from the low noise of the velocity reconstruction and thus the high correlation coefficient $r$ that can exceed $99\%$ for a dense survey like LSST. For LSST and CMB S4 combined we find that one can reach $\sigma_{\fnl} \sim 0.5$, a potential two sigma exclusion of the multifield bound, and an improvement factor of about a factor of $3$ with respect to the galaxy survey alone (assuming no internal sample variance cancellation of galaxies). This forecast includes marginalization over all relevant parameters and realistic photo-z errors, but neglects catastrophic redshift errors. If one can mass bin galaxies, the improvement factor strongly depends on the biases and densities of the mass binned galaxies, but a simplified forecast using halos shows that a factor of two improvement is still realistic in a relevant range of halo masses. Our method is important, as we obtain significant sensitivity improvements on the $\fnl$ parameter, which come at no additional experimental costs. Constraining local non-Gaussianities below the multi-field inflation threshold is one of the key science motivations for upcoming large-scale galaxy surveys, and kSZ tomography helps significantly to achieve this goal.

In this paper we have used a simplified 3-dimensional box geometry, which was ideal to illustrate the power and properties of the method without evaluating complicated geometric projection integrals. A fully realistic treatment requires spherical coordinates and will appear in~\cite{kszfnlholst}, in a combined analysis also including CMB lensing for additional sample variance cancellation from transverse modes as in~\cite{Schmittfull:2017ffw}. We will also investigate the contribution of the primary CMB and ISW effects on the ``effective velocity'' discussed in~\cite{Terrana2016}. Another interesting direction would be to include marginalization over parameters of the electron profile entering in $P_{ge}$, although we do not expect the $\fnl$ forecast to depend on this significantly. Furthermore, a straightforward extension of this work can provide constraints on higher-order non-Gaussianities such as the $g_{NL}$ model \cite{Smith:2011ub}. Finally the simple quadratic estimator for kSZ velocity reconstruction may suffer from shortcomings due to the highly non-linear matter field at small scales in a similar way as the quadratic lensing potential estimator from the CMB, which warrants further investigation. We believe that kSZ tomography will be an important component to push $\fnl$ constraints below the theoretical target of $\fnl\simeq1$.

\section*{Acknowledgments}

Research at Perimeter Institute is supported by the Government of Canada
through Industry Canada and by the Province of Ontario through the Ministry of Research \& Innovation.
MSM is grateful to Perimeter for supporting visits during which this work was carried out.
KMS was supported by an NSERC Discovery Grant and an Ontario Early
Researcher Award. MCJ is supported by the National Science and Engineering Research Council through a Discovery grant. SF was funded by the Miller Fellowship at the University of California, Berkeley.
We thank Pat McDonald, Neal Dalal, Jo Dunkley, Emmanuel Schaan, Marcel Schmittfull, Uros Seljak, David Spergel and Martin White for useful discussions, and Emanuela Dimastrogiovanni for collaboration during the early stages of this work.

\bibliography{ksz_fnl_paper}

\appendix

\section{Halo model power spectra}
\label{app:halomodel}

In this appendix we collect the mass binned halo model equations used in our forecast. The halo mass function and HOD which we use are described in App. B of~\cite{kszbispectrumpaper}. The mass binned halo power spectra for mass bins $i$ and $j$ are
\begin{eqnarray}
P_{hh,ij} (k,z) &=& \delta_{ij} P_{hh,i}^{1h}(k,z) + P_{hh,ij}^{2h}(k,z)\\
P_{hh,i}^{1h}(k,z) &=& \frac{1}{n_{h,i}}\\
P_{hh,ij}^{2h}(k,z) &=& b_{h,i}(z) \ b_{h,j}(z) \ P_{\rm lin}(k,z) 
\end{eqnarray}
where the 1-halo term only arises for the diagonal case $i=j$ (all mass bins are defined non-overlapping). Here we have defined the mean halo bias in the bin
\be
b_{h,i}(z) = \frac{1}{n_{h,i}} \int_{m_{i,{\rm min}}}^{m_{i,{\rm max}}} dm \ n(m,z)  b_h(m,z) 
\ee
and the mass binned halo number density
\be
n_{h,i}(z) = \int_{m_{i,{\rm min}}}^{m_{i,{\rm max}}} dm \ n(m,z).
\ee
The mass binned power spectra for galaxies are
\begin{eqnarray}
P_{gg,ij} (k,z) &=& \delta_{ij} P_{gg,i}^{1h}(k,z) +  P_{gg,ij}^{2h}(k,z) + \frac{\delta_{ij}}{n_{g,i}(z)}\\
P_{gg,i}^{1h}(k,z) &=& \int_{m_{i,{\rm min}}}^{m_{i,{\rm max}}} dm \ \frac{n(m,z)}{n_{g,i}^2} \left[ 2 \langle N_c(m) N_s(m) \rangle |u_s(k|m,z)| + \langle N_s(m) (N_s(m)-1) \rangle |u_s(k|m,z)|^2 \right] \\
P_{gg,ij}^{2h}(k,z) &=& P^{\rm lin} (k,z) \left[ \int_{m_{i,{\rm min}}}^{m_{i,{\rm max}}} dm \ n(m,z)  b_h(m,z) \frac{\langle N_c(m) \rangle + \langle N_s(m) \rangle u_s(k|m,z)}{n_{g,i}} \right] \\
&&\times \left[ \int_{m_{j,{\rm min}}}^{m_{j,{\rm max}}} dm \ n(m,z)  b_h(m,z) \frac{\langle N_c(m) \rangle + \langle N_s(m) \rangle u_s(k|m,z)}{n_{g,j}} \right] 
\end{eqnarray}
with galaxy density
\be
n_{g,i}(z) =  \int_{m_{i,{\rm min}}}^{m_{i,{\rm max}}} dm \ n(m,z) \left( \langle N_c(m) \rangle + \langle N_s(m)  \rangle \right).
\ee
and galaxy bias
\begin{equation}
b_{g,i} (z) = \frac{1}{n_{g,i}} \int_{m_{i,{\rm min}}}^{m_{i,{\rm max}}} dm \ n(m,z)  b_h(m,z) \left(\langle N_c(m) \rangle + \langle N_s(m) \rangle \right).
\end{equation}
For our fiducial model, we assume that the normalized fourier transform of the satellite galaxy profile $u_s(k|m,z)$ is NFW, tracing the dark matter. We also need the mass binned cross power of halos and galaxies with electrons to calculate $N_{vv}$. For halos we obtain
\begin{eqnarray}
P_{he,i} (k,z) &=& P_{he,i}^{1h}(k,z) + P_{he,i}^{2h}(k,z) \\
P_{he,i}^{1h}(k,z) &=& \int_{m_{i,{\rm min}}}^{m_{i,{\rm max}}} dm \ n(m,z) \left( \frac{m}{\rho_m} \right) u_e(k|m,z) \frac{1}{n_{h,i}} \label{eq:phe_1h} \\
P_{he,i}^{2h}(k,z) &=& P^{\rm lin} (k) \ b_{h,i}(z) \left[ \int_{-\infty}^\infty dm \ n(m,z) \left( \frac{m}{\rho_m} \right) b_h(m,z) u_e(k|m,z) \right]  \label{eq:phe_2h}
\end{eqnarray}
For the galaxy-electron cross power we get 
\begin{eqnarray}
P_{ge,i} (k,z) &=& P_{ge,i}^{1h}(k,z) + P_{ge,i}^{2h}(k,z) \\
P_{ge,i}^{1h}(k,z) &=& \int_{m_{i,{\rm min}}}^{m_{i,{\rm max}}} dm \ n(m,z) \left( \frac{m}{\rho_m} \right) u_e(k|m,z) \frac{\langle N_c(m) \rangle + \langle N_s(m) \rangle u_s(k|m,z)}{n_{g,i}} \label{eq:pge_1h} \\
P_{ge,i}^{2h}(k,z) &=& P^{\rm lin} (k) \left[ \int_{m_{i,{\rm min}}}^{m_{i,{\rm max}}} dm \ n(m,z)  b_h(m,z) \frac{\langle N_c(m) \rangle + \langle N_s(m) \rangle u_s(k|m,z)}{n_{g,i}} \right] \nonumber \\ 
&& \times \left[ \int_{-\infty}^\infty dm \ n(m,z) \left( \frac{m}{\rho_m} \right) b_h(m,z) u_e(k|m,z) \right]  \label{eq:pge_2h}
\end{eqnarray}
The normalized fourier transform of the electron distribution in halos $u_e(k|m,z)$ is given by the "AGN" model of Ref.~\cite{kszbispectrumpaper}.

\end{document}